\documentstyle[11pt,paspconf,epsf]{article}
\begin{document}

\title{Eigenmode Analysis of Radial Velocities}
\author{Y. Hoffman}
\affil{Racah Inst. of Physics, Hebrew University, Jerusalem 91904,
        Israel}

\begin{abstract}

Radial velocity surveys are examined in terms of eigenmode analysis 
within the framework of CDM-like family of models. Rich surveys such as 
MARK III and SFI, which consist of more than $10^{3}$ radial velocities, 
are found to have a few tens of modes that are  not noise dominated. 
Poor surveys, which have only a few tens of radial velocities, are   
noise dominated across the eigenmode  spectrum. In particular, the bulk 
velocity of such surveys has been found to be dominated by the more 
noisy modes. The   MARK III and SFI are well fitted by a tilted   flat CDM 
model found by   a maximum likelihood analysis and a $\chi^{2}$ statistics . 
However, a mode-by-mode inspection shows that a substantial 
fraction of the modes lie outside the $90\% $ confidence level. This 
implies that although globally the CDM-like family of models seems 
to be consistent with radial velocity surveys, in detail it does not. 
This might indicate a need for a revised power spectrum or for 
some non-trivial biasing scheme.

\end{abstract}


\section{Introduction}
The statistical analysis of surveys of radial velocities plays a major 
role in the study of the large scale structure.  Broadly speaking the 
analysis focuses on the estimation of the cosmological parameters and 
the reconstruction of the local cosmography.  Radial velocities 
surveys are dominated by incomplete and anisotropic sky coverage, 
inhomogeneous, sparse sampling and distance measurements errors that 
are often larger than the individual velocities.  Such surveys do not 
easily yield themselves to a statistical analysis and provide a real 
challenge for a proper analysis.  The history of the field is 
therefore rich with controversies about the interpretation and 
cosmological implications of different surveys and seemingly 
conflicting results.

A prime goal of a statistical analysis is to provide   a tool 
for confronting theoretical models with the data.  This calls for a 
formalism for presenting models and data in the same language, a 
problem closely related to the problem of the functional 
representation.  From the theoretical point of view, the choice of the
representation is dictated by the symmetries of the theory.  
The cosmological principle makes the Fourier plane waves and the 
spherical Bessel/harmonics the natural choice.  However typical 
astronomical observations break these symmetries, as the data is neither 
homogeneous nor isotropic.  A better representation should reflect both 
the basic underlying theory and the particularities of the data.  Here 
we follow the standard eigenmode analysis, also 
known as principal components analysis (PCA) and the Karhunen-Loeve 
transform.  This, or its slightly modified version of signal-to-noise 
eigenmodes, was suggested before as a method of analyzing 
redshift surveys (Vogeley and Szalay 1996 and references therein). 
Previous applications focused mostly on parameters estimation .  
Here we extend the 
method and use it as a general tool for understanding the nature of 
the data, its noise structure and information content.  Dealing with 
velocity surveys the determination of bulk velocities of a given 
survey and its relation to the underlying 
velocity field is   revisited.  The PCA is used here to address 
the problem of the power spectrum determination.

\section{Eigenmode Analysis: Radial Velocities and Bulk Velocities}


Consider a data base of radial velocities $\{ u_{i}\}_{i=1,\ldots,N}$, 
where 
\begin{equation}
u_{i} = {\bf  v}
({\bf r}_{i})  \cdot \hat {\bf r}_{i} 
+\epsilon_{i},
	\label{ui}
\end{equation}
${\bf  v}
$ is the three dimensional velocity, ${\bf r}
_{i}$ is the 
position of the i-th data point  and $\epsilon_{i}$ is the statistical 
error associated with the i-th radial velocity. The assumption made 
here is of   a   cosmological model that  well 
describes the data, that systematic errors have been properly dealt 
with and that the statistical errors are well understood. 
The data auto-covariance matrix  is then  written as:
\begin{equation}
R_{ij} \equiv
\Bigl < u_i u_j  \Bigr > =   \hat {\bf r}_j \Bigl < {\bf  v}
({\bf r}_i) {\bf  v} ({\bf r}_j)  \Bigr > 
   \hat {\bf r}_j   + \sigma{^2_{ij}}.
\label{eq:Rij}
\end{equation}
(Here $\Bigl <  \ldots  \Bigr >$ denotes an ensemble average.)
The last term is the error covariance matrix.
The velocity covariance tensor that enters this equation was derived 
by  G\'orski (1988, see also Zaroubi, Hoffman and Dekel 
1999) and it depends on the power spectrum and the cosmological 
parameters.

The eigenmodes of the data covariance matrix provides a natural 
representation of the data:
\begin{equation}
R {\bf\eta}^{(i)}€ =\ \lambda_{i}€{\bf\eta}^{(i)}
\label{eq:eigenvec}
\end{equation}
The set of $N$ eigenmodes   $\{ {\bf \eta}^{(i)}\}$ constitutes an 
orthonormal basis and the eigenvalues $\lambda_{i}$  are arranged in 
decreasing order. A new representation of the data   is given by:
\begin{equation}
\tilde a_{i}€= \eta{^{(i)}_{j}} \  u_{j}
\label{eq:aiui}
\end{equation}
This provides a 
statistical orthogonal representation, namely:
\begin{equation}
\bigl\langle \tilde a_{i}€\tilde a_{j}€\bigr\rangle \ = 
 \ \lambda_{i}€\delta_{ij}€
\label{eq:tildeaiaj}
\end{equation}
The normalized transformed variables are defined here by:
\begin{equation}
a_{i}€ = {\tilde a_{i}€ \over \sqrt{\lambda_{i}} }
\label{eq:defai}
\end{equation}
Eq. \ref {eq:tildeaiaj} is written now as:
\begin{equation}
\bigl\langle \tilde a_{i}€\tilde a_{j}€\bigr\rangle \ = 
 \ \delta_{ij}€
\label{eq:aiaj}
\end{equation}
Note that as the modes are statistically independent one can measure 
the $\chi^{2}$ of a given mode, independently of all other modes:
\begin{equation}
\chi{^{2}_{i}}=a{^{2}_{i}}
\label{eq:chi2i}
\end{equation}
For normally distributed errors and a Gaussian random velocity field 
the $a_{i}$'s are normally distributed with zero mean and a variance 
of unity.

Velocity surveys are often analyzed in terms of their bulk flows, 
namely fitting the velocity field by a single constant velocity 
vector, ignoring any possible correlations of the underlying field.  
There are a variety of ways of defining the bulk velocity of a given 
survey and here we adopt the  Kaiser (1988) algorithm which evaluates an 
error weighted bulk velocity. Thus, the 
full complexity of the underlying field of its  N degrees of freedom 
is compressed into three parameters only.  It is 
often argued that this data compression enables the extraction of more 
statistically robust quantities from the data, thus providing better 
constraints on the models. The bulk velocity properties of a survey 
is studied here within the PCA formalism.

The bulk velocity (${\bf B}$) of a survey is defined by means of a linear 
operator ($L$),  ${\bf B} = L {\bf u}$ (see Kaiser 1988 for the 
formal derivation). ${\bf B}$ is expanded here by
\begin{equation}
 {\bf B}  = \sum_{i}€a_{i}€{\bf B}^{(i)} = \sum_{i}€a_{i}€
             \sqrt{\lambda_{i}} L {\bf\eta}^{(i)},  
\label{eq:bi}
\end{equation}
where ${\bf B}^{(i)}$ is the bulk 
velocity associated with the i-th mode.
The bulk velocity covariance matrix is:
\begin{equation}
 \bigl< B_{\alpha}€B_{\beta}\bigr> = \sum_{i}{B{^{(i)}_{\alpha}} 
B{^{(i)}_{\beta}}} 
\label{eq:bbcov}
\end{equation}
In the case of an anisotropic sampling the bulk velocity covariance 
matrix is anisotropic as well.
In the limit of  a perfect survey (isotropic, no errors, dense, homogeneous) the 
eigenvectors  are the spherical harmonics and Bessel functions.  
In such a case and assuming the data to lie on a  thin shell, 
one expects:
\begin{equation}
 B{_{\alpha}^{(i)} } = 0 \ {\rm for} \  i \neq {2,3,4} 
\label{eq:B123}
\end{equation}

\section{Observations}

The problem to be addressed here is the quality and expected 
significance of a survey given an assumed model, {\it i.e.} power 
spectrum, of the underlying velocity field.  It follows that here one 
is more interested in the sampling, sky coverage and the errors   
then in the actual numerical value of the data points.  Four data sets 
are studied here: The MARK III catalog (Willick {\it et al.} 1995), 
SFI (da Costa {\it et al.} 1996), LP10K (Willick 1999) and the 
nearby Type Ia supernova (hereafter SN; Riess 1999) .  The first two 
data sets consist of more than 1000 radial velocities, and are 
considered as rich catalogs, where the other two have velocities of 
only 15 Abell clusters (LP10K) and 44 Type Ia supernova (SN) and are 
considered here as poor catalogs.  The analysis has been applied to a 
wide range of CDM-like models but only two models are explicitly 
presented here.  One is a flat tilted CDM model ($n=0.8, \ h=0.75, \ 
\Omega_{0}=1$, where $n$ is the power index, $h$ is Hubble's constant 
in units of $100 Mpc^{-1} km s^{-1}$ and $\Omega_{0}$ is the density 
parameter), and the flat $\Lambda$-CDM model with $n=1, \ h=1, \ 
\Omega_{0}=0.4$.  Both models are COBE normalized.  The tilted model 
is the most probable CDM-like model for the MARK III data (Zaroubi 
{\it et al.} 1997), and is very close to the model favored by the 
SFI (Freudling  {\it et al.} 1999).  The other model is the 
currently most popular model obeying the age and geometrical 
constraints.  The results obtained for the two models are basically 
very similar.  The strategy followed here is to compute the eigenmodes 
and eigenvalues of a given survey and model with and without the 
noise.  The comparison of these reveals how many independent modes are 
signal or noise dominated.  It can also help assessing the degree to 
which the bulk velocity of the sample reflects the underlying velocity 
field or the observational errors.

For all data sets the noise-free eigenmode spectrum follows an 
approximate power law behavior over most of the range of modes.  The 
addition of noise breaks this power law decline, and causes a 
flattening of the spectrum. The transition from one regime to the 
other marks the transition from the signal to noise dominated regimes. 
There is a striking  difference between the rich surveys 
(MARK III and SFI, Fig. 1) and the poor surveys (SN and LP10K, Fig. 2). The formers 
exhibit a clear break, with some 10 modes or so that are virtually 
unaffected by the noise and a few tens of modes  that are not noise 
dominated. In the poor samples, on the other hand, all modes 
are noise dominated! This happens for a wide spread of acceptable 
CDM-like models, both with COBE and clusters normalization. 

\begin{figure} 
\plottwo{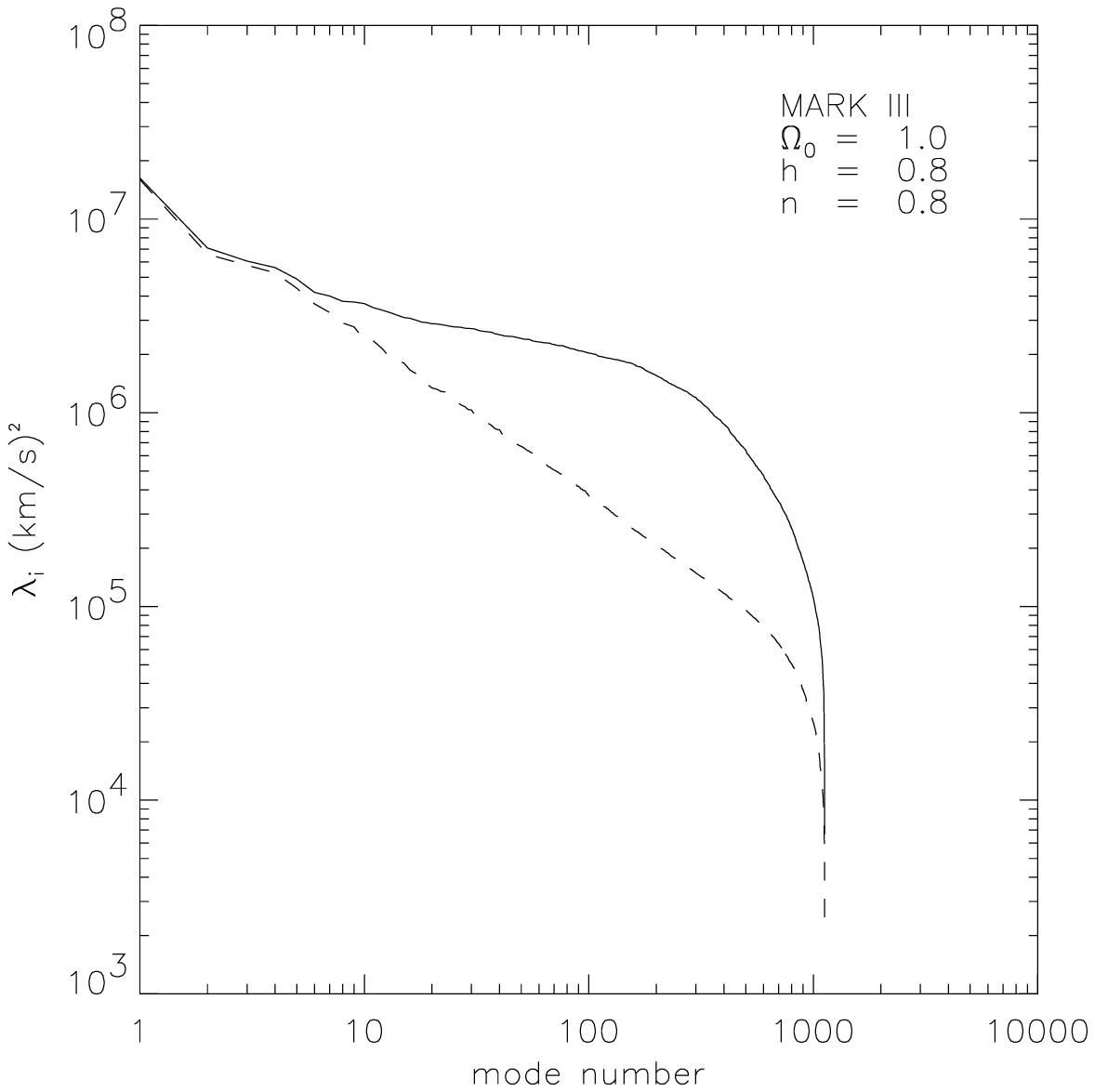}{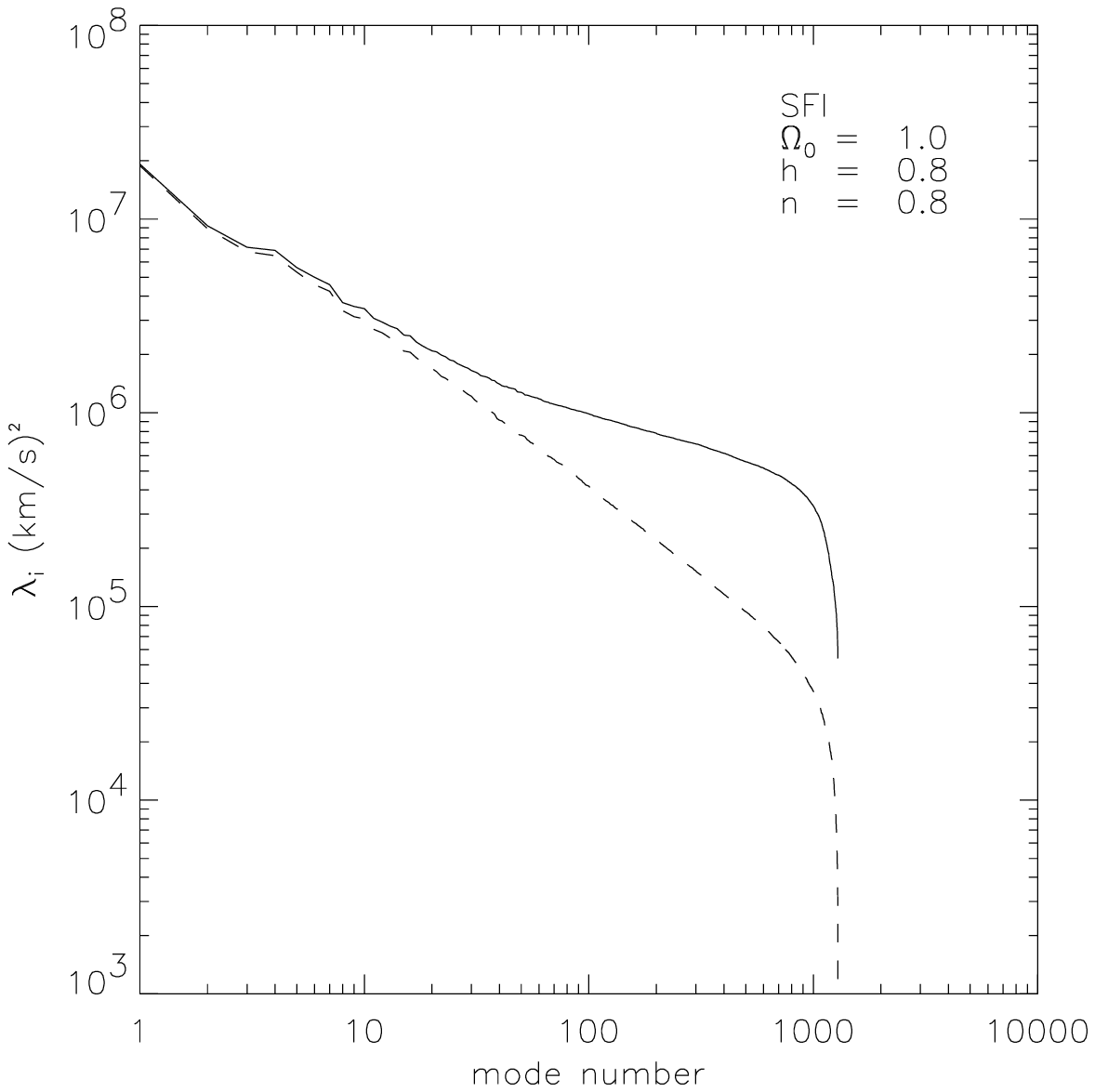}
\caption{The eigenvalue spectrum is plotted against the mode number 
for the MARK III (left) and SFI (right) data. The solid line is the 
spectrum of the noise-free covariance matrix and the dashed line 
corresponds to the full (signal+noise) covariance matrix. This is 
calculated for the tilted-CDM model.  }
\label{fig:m3sfi-spectrum}
\end{figure}
\begin{figure} 
\plottwo{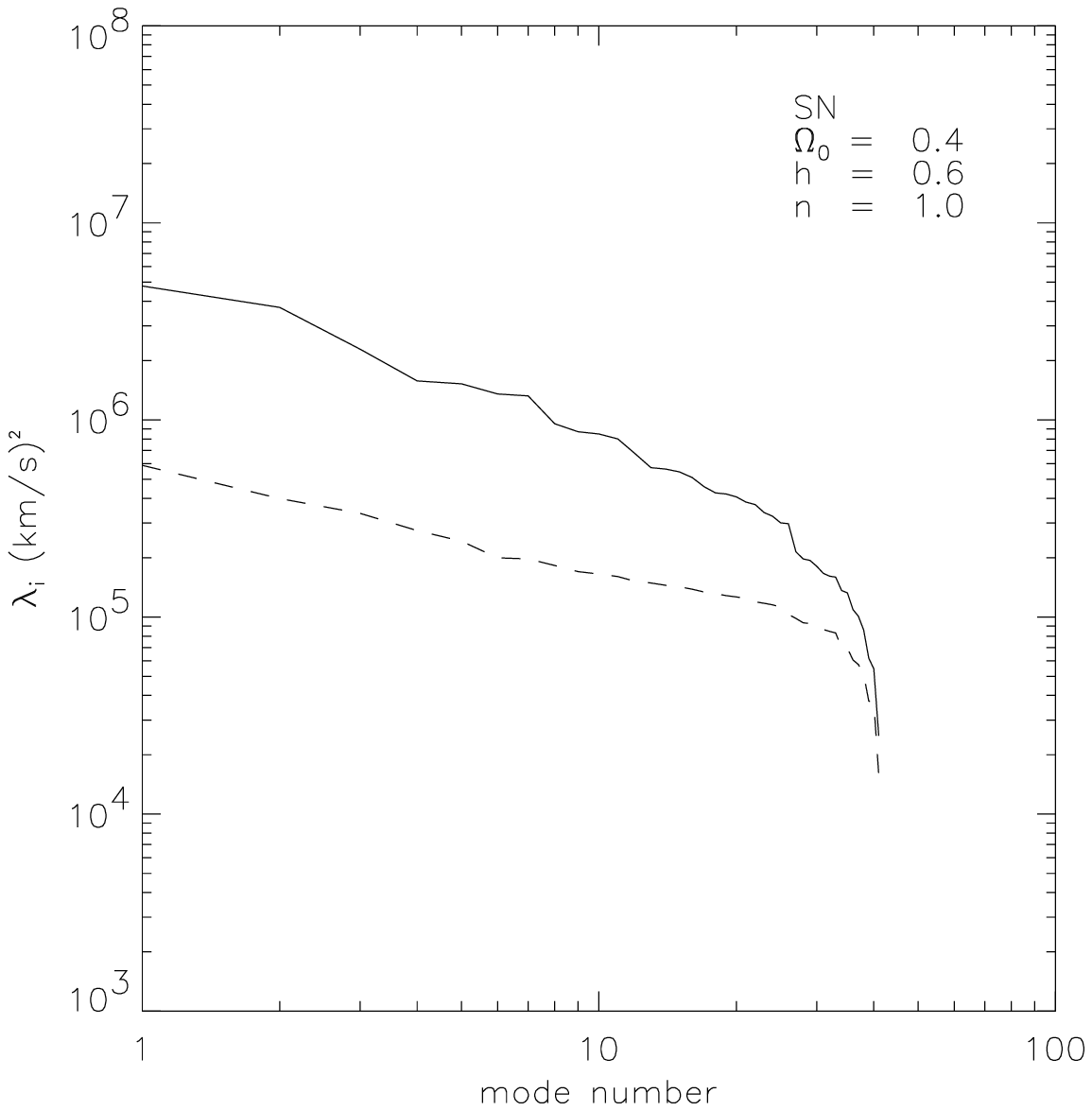}{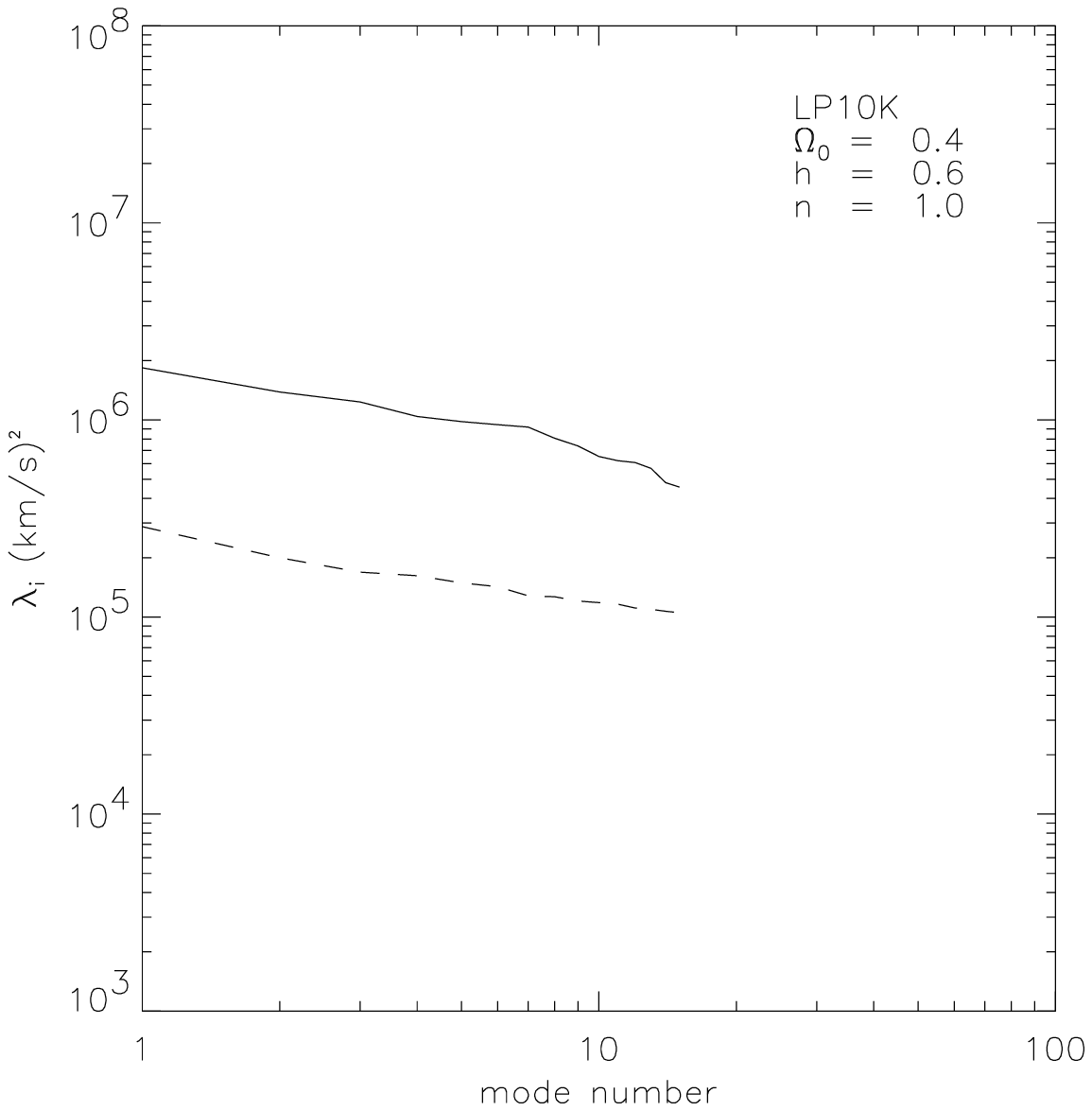}
\caption{The eigenvalue spectrum is plotted against the mode number 
for the SNI (left) and LP10K (right) data. This is 
calculated for the  $|ambda$-CDM model. Same notations as in 
Fig. 1. }
\label{fig:SNLP10K-spectrum}
\end{figure}

Next, the bulk velocities of the spectrum of eigenvalues, $B^{(i)}$, 
is calculated.  For an ideal survey only the first few modes are 
expected to be significant and the rest of the modes should have very 
small bulk velocities.  MARK III and SFI indeed show such a behavior, 
namely the $B^{(i)}$ of the first few modes lie significantly above the 
noise level (Fig.  3).  For the poor samples, SN and LP10K, an 
opposite trend is found as $B^{i}$ does not decline, or even grows, 
with the mode number, namely the more dominant by the noise a mode is 
the higher is its $B^{i}$ (Fig.  4).  It follows that the sample bulk velocity 
of the rich surveys indeed reflects the underlying 
velocity field (convolved with the sample window function).  In the 
case of the poor samples the bulk velocity is dominated by the noise.  
It should be stated   that the strong conclusions expressed 
here are valid only within the framework   of CDM-like 
comsogonies.

\begin{figure} 
\plottwo{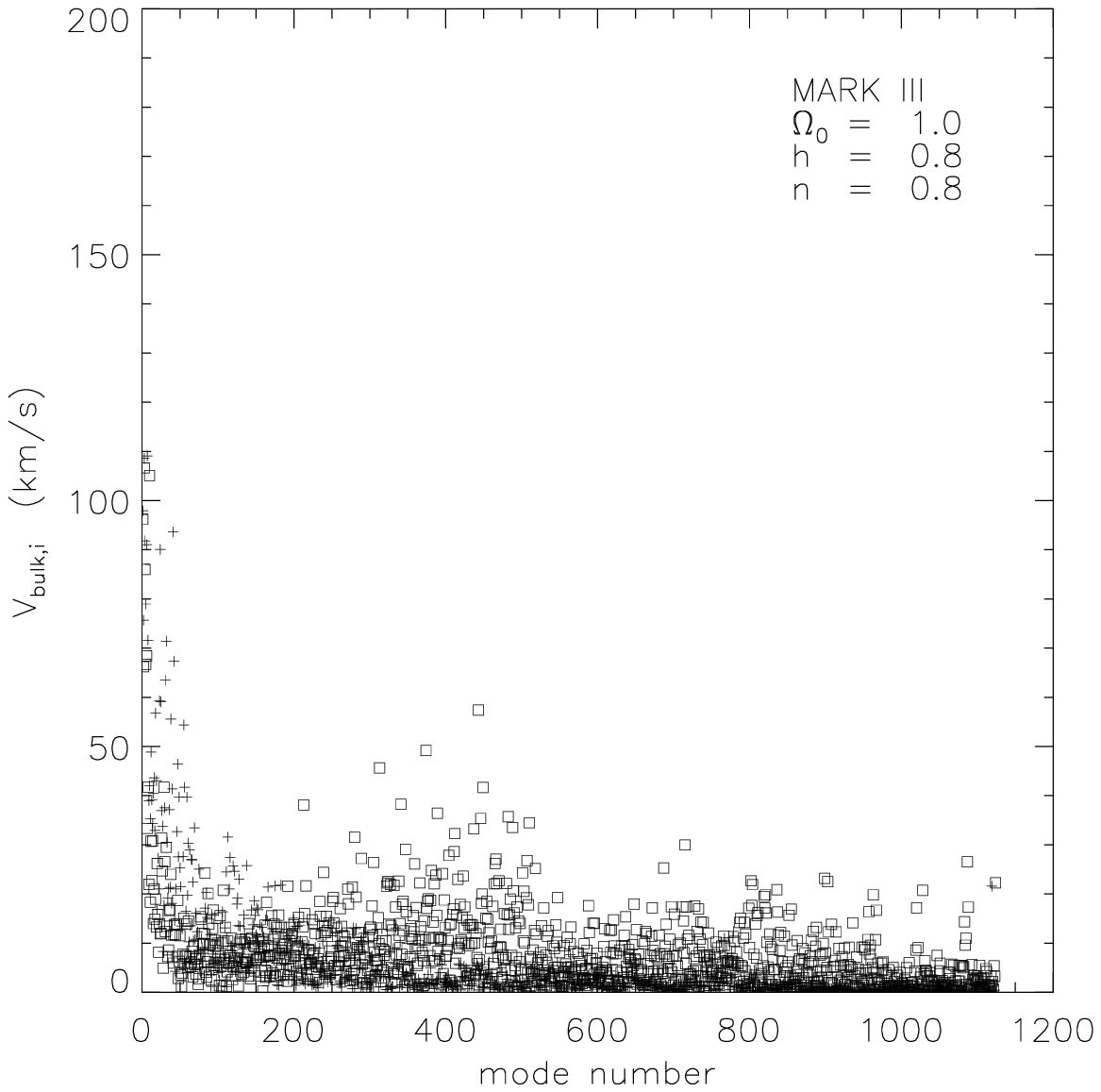}{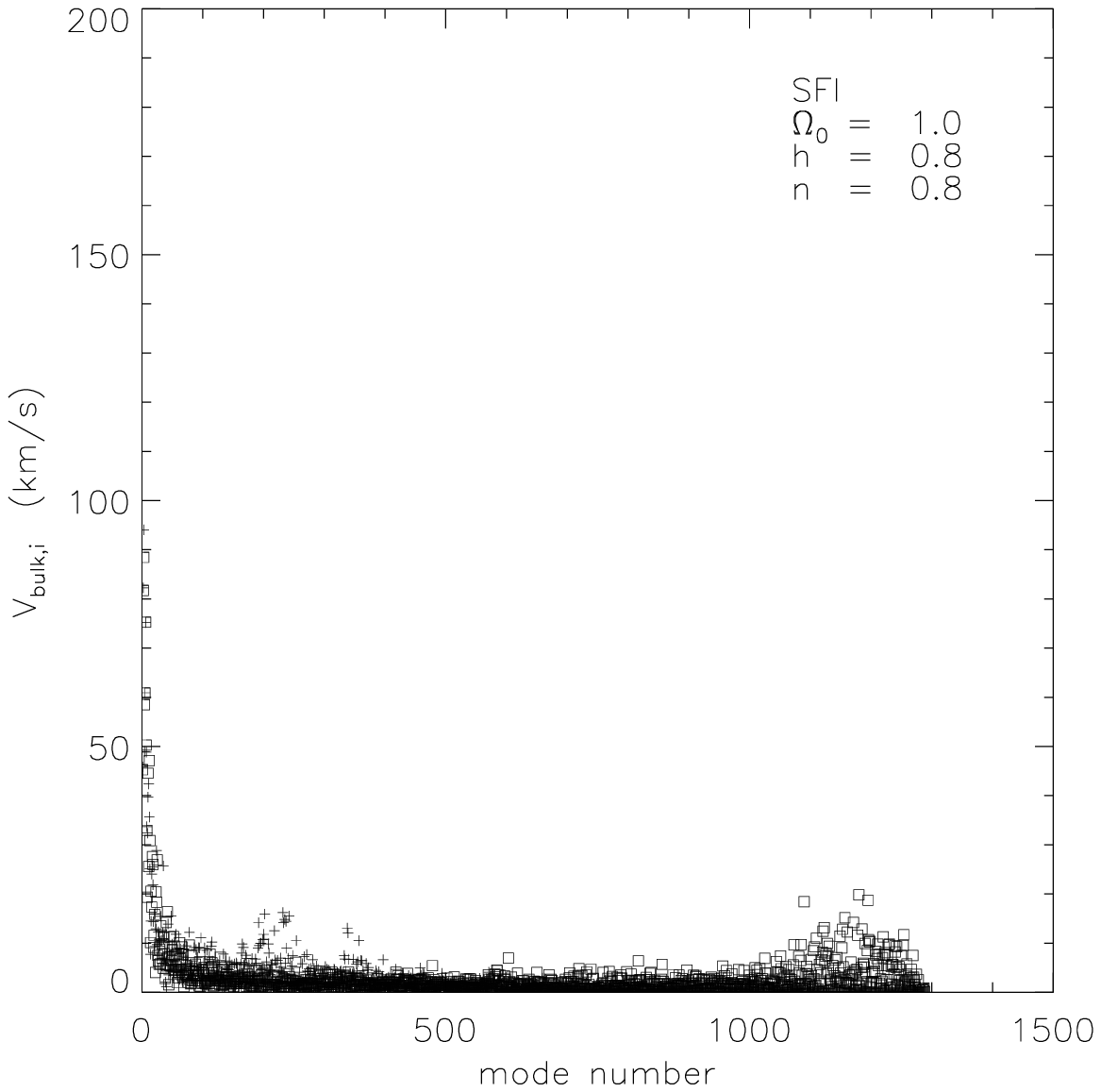}
\caption{The bulk velocities  spectrum is plotted against the mode number 
for the MARK III (left) and SFI (right) data. The crosses correspond 
to the 
spectrum of the signal covariance matrix and the squares 
to the full (signal+noise) covariance matrix. This is 
calculated for the tilted-CDM model.  }
\label{fig:m3sfi-Bi}
\end{figure}
\begin{figure} 
\plottwo{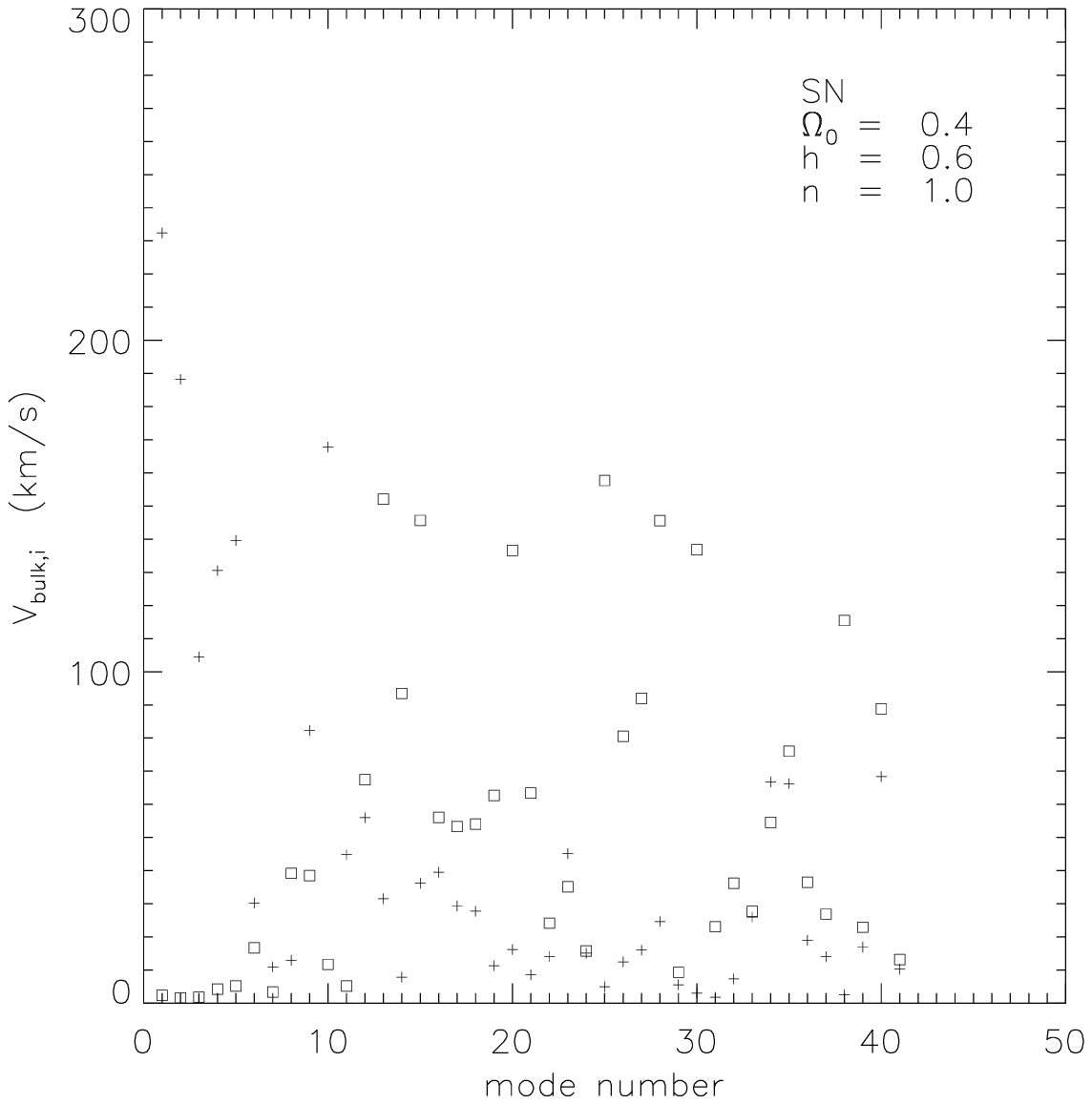}{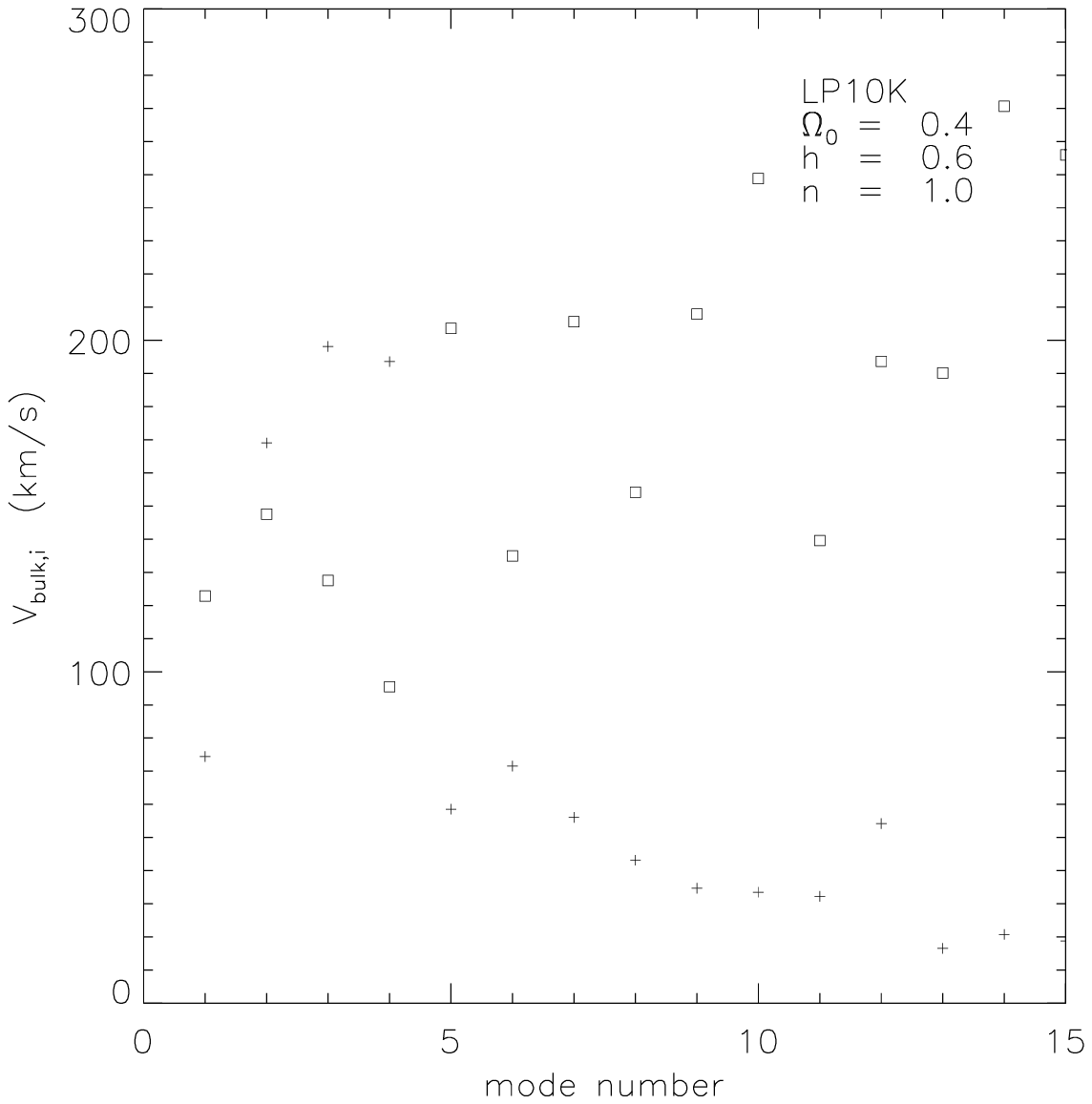}
\caption{The bulk velocities  spectrum is plotted against the mode number 
for the SN (left) and LP10K (right) data.   This is 
calculated for the  $\lambda$-CDM model. Same notations as in 
Fig. 3. }
\label{fig:SNLP10K-Bi}
\end{figure}

\section{Power Spectrum}                       
\label{sec:ps}
The optimal way of estimating the values of the cosmological 
parameters from a given survey is by performing a maximum likelihood 
analysis of the data given a range of models (Zaroubi {\it et al.} 
1995).  The maximum likelihood analysis does 
not provide an absolute measure of the quality of the parameter 
estimation, but rather finds the most probable model given the data 
and the assumed parameter space.  A measure of the goodness of fit is 
provided by the $\chi^{2}/{\rm d.o.f}$.  However, it is often the case 
that the best model of a given parameter space and a given data set 
fits poorly on some scales and better on some others, small {\it vs\ } 
large scales say.  This can result in a `conspiracy', yielding an 
adequate $\chi^{2}$ even for the `wrong' model.    The PCA   
which projects the data into statistically 
independent normal variables   enables the analysis of 
the data on a mode by mode basis and check the goodness of fit across 
the spectrum of the modes.  One recalls here that the $\chi^{2}$ of a 
given mode is simply $a{^{2}_{i}}$ and the cumulative $\chi^{2}$ is
\begin{equation}
\chi{^{2}_{M}}={\sum_{i=1}^{M} { a{^{2}_{i}} } \over M }.
	\label{eq:chim}
\end{equation}

This combined PCA and $\chi^{2}$ analysis is applied here to the SFI 
and MARK III catalogs.  The smaller data sets (SN and LP10K) are not 
discriminative enough to enable such a study.)  Fig.  5 shows the 
cumulative $\chi{^{2}_{M}}$ of the MARK III and SFI surveys (assuming 
the tilted-CDM model).  On top of these plots the lower and upper 
$90\% $ confidence levels are plotted.
For the the MARK III the total $\chi^{2}/d.o.f=1.02$ is well within 
the $90\% $ limit and therefore seems to be very consistent with the data.  
However, Fig.  5 shows that over most of the mode number range the 
$\chi{^{2}_{M}}$ lies outside the $90\% $
confidence band.  Actually from approximately the 100th mode to the 
last one there is a monotonic increase of $\chi{^{2}_{M}}$.  A similar 
trend is also exhibited by the SFI data.  To check the constraining 
power of the PCA/$\chi^{2}$ test it has been applied to the mock MARK 
III catalog of Kolatt {\it et al.} (1996).  The cumulative 
$\chi{^{2}_{M}}$ is found to be fully consistent with the assumed 
model (figures are not shown here).  Thus, a systematic 
inconsistency of the best CDM-like model with the data is found here 
that  possibly 
suggests a fundamental problem to the CDM paradigm.  
A more detailed analysis is to be presented elsewhere.
\begin{figure} 
\plottwo{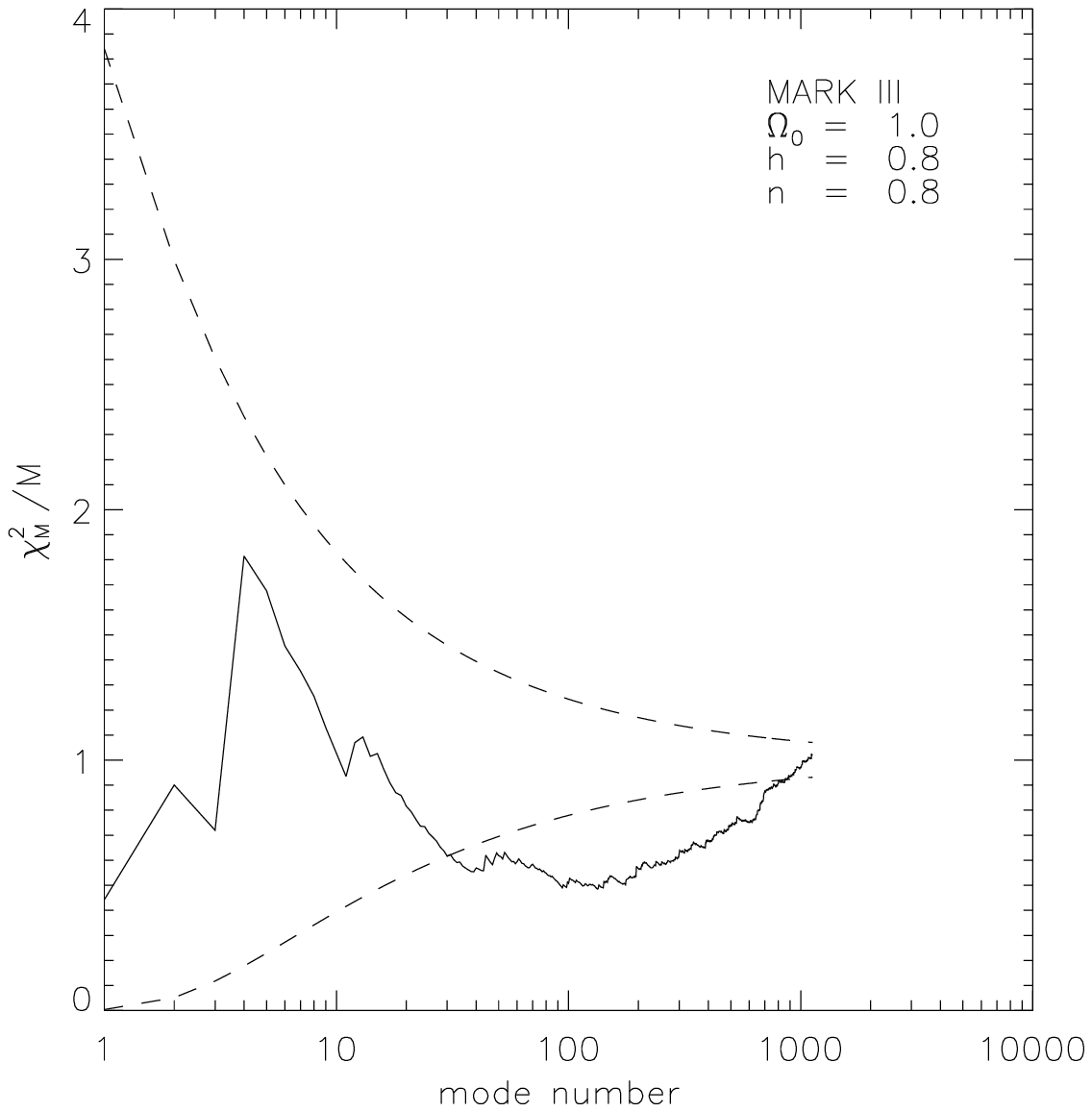}{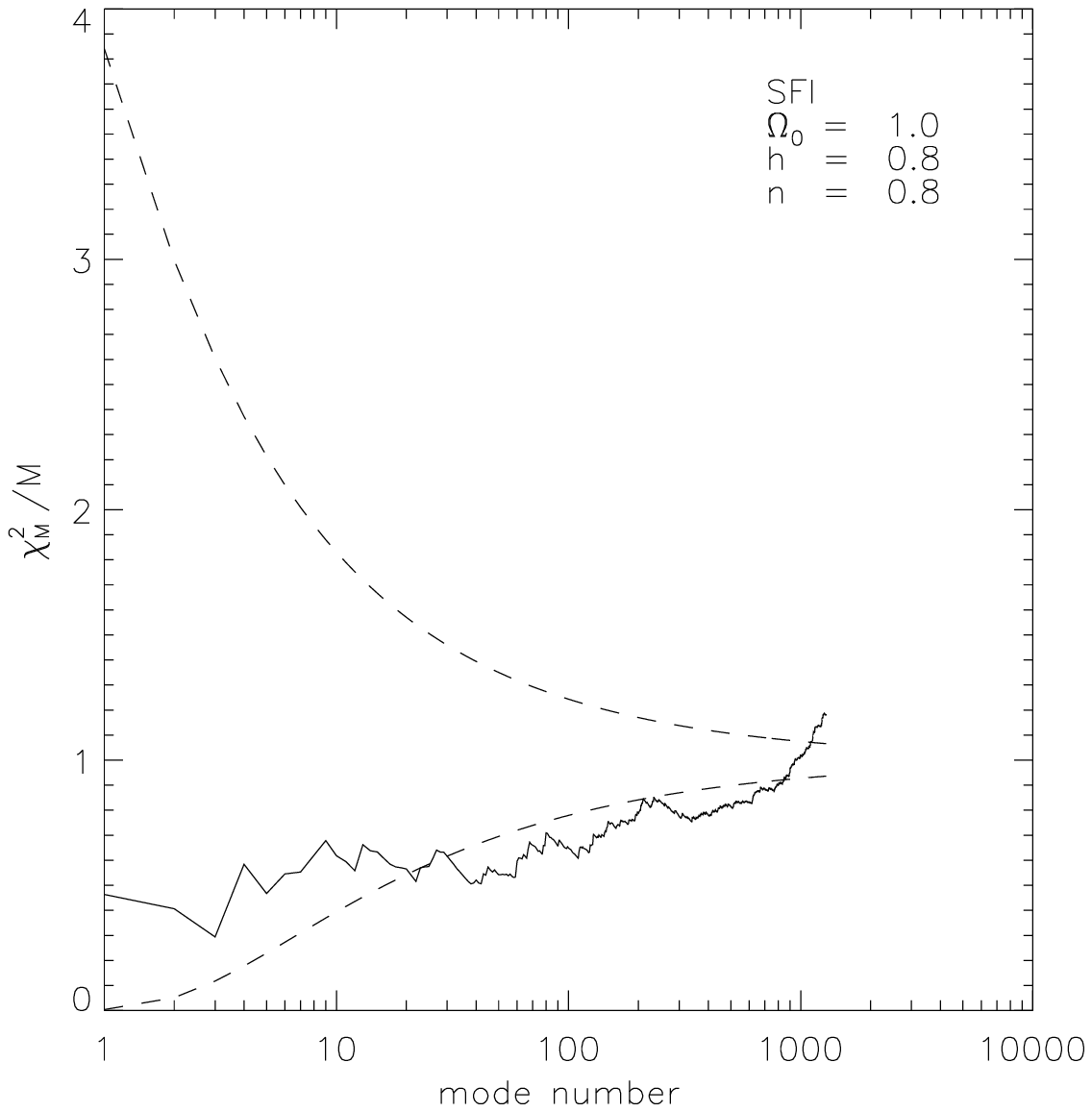}
\caption{Cumulative $\chi^{2}$ analysis of the MARK III (left) and SFI (right) 
data for  a flat tilted 
($n=0.8$)  CDM model. The upper and lower $5\% $ and $95\% $ 
confidence levels are plotted for reference. }
\label{fig:m3-chi}
\end{figure}

\section{Discussion}

The analysis presented here consists of two parts. First, the 
constraining power of radial velocities surveys has been examined 
within the CDM-like family of models. Using PCA the structure of the 
expected data has been considered, rather then the actual numerical 
value of the data. The analysis reveals that rich surveys such 
as MARK III and SFI, have a few tens of modes that are not noise 
dominated, and hence are expected to reflect the underlying velocity 
field. Poor samples such as LP10K, SN and most probably all other surveys 
that consist of a few tens of objects are noise dominated. 
Not even a single eigenmode is signal dominated for such surveys, and 
the bulk velocity is dominated by the more noisy, and less 
significant, modes. This does not imply that such surveys are of no 
use in cosmology, but that they should be analyzed with great care. 
Direct reconstruction methods  might be completely noise 
dominated and might be very misleading. Indirect methods such as 
Wiener filtering and maximum entropy should be useful in analyzing 
such data. A note of cautious is due here. The statements made here are 
valid only within the framework of the standard cosmogony of 
CDM-like family of models.

Having convinced ourselves that surveys such as MARK III and SFI are 
powerful enough to constrain the CDM-like models, the consistency of 
these surveys with the models has been examined in detail.  A 
mode-by-mode inspection finds significant discrepancies with the 
spectral behavior predicted by the `best' model found by the maximum 
likelihood analysis and a global $\chi^{2}€$ analysis.  It seems that 
the overall agreement is obtained by some `conspiracy', where the 
combination of the independent modes yields a reasonable $\chi^{2}€$.  
 This implies a gross disagreement of the most 
favorable cosmological model with the velocity data, or the need to 
invoke some non-trivial biasing.

 PCA can  also play a very important role in designing and planning 
 new surveys. PCA is based on analyzing the data covariance matrix, 
 which expresses the statistical properties of the data rather then 
 its actual numerical values. It follows that it can be applied 
 before a survey is done, and therefore can be used  to design it.
 By studying the spectrum and structure of the eigenmodes of a survey 
 of given geometry and depth and expected errors the constraining 
 power of a survey can be properly evaluated in its planning phase.

\acknowledgments
I would like to thank my collaborator on this project and many others, Saleem 
Zaroubi. I have benefited from many interesting discussions with 
Enzo Branchini, Avishai dekel and Ofer Lahav. This research has been 
partially supported by a Binational Science Foundation grant 94-00185 and 
an Israel Science Foundation grant 103/98.


\begin{references}


\reference da Costa, L.N., Freudling, W., Wegner, G., Giovanelli, R.,
  Haynes, M.P., \& Salzer, J.J. 1996, \apjl,  468 , L5

\reference Freudling, W., Zehavi, I., da Costa, L.N., Dekel, A.,
   Eldar, A., Giovanelli, R., Haynes, M.P., Salzer, J.J., Wegner, G.,
   \& Zaroubi, S. 1999, preprint (astro-ph/9904118)


\reference G\'orski, K.M. 1988, \apj, 332, L7


\reference Kolatt T., Dekel A., Ganon G. \& Willick J. 1996, ApJ, 458, 419

\reference Riess A.G. 1999, These proceedings (astro-ph/9908237)

\reference Vogeley M.S. \& Szalay A.S. 1996, \apj, 465, 34

\reference Willick, J.A. 1999, These proceedings (astro-ph/9909004)

\reference Willick, J.A., Courteau, S., Faber, S.M., Burstein, D.,
  \& Dekel. A. 1995, ApJ, {446}, 12


\reference Zaroubi, S., Hoffman, Y.,  Fisher, K.B., \& S. Lahav, O. 1995, ApJ,
{449}, 446

\reference Zaroubi, S., Hoffman, Y. \& Dekel, A. 1999, \apj, 520, 413

\reference Zaroubi, S., Zehavi, I., Dekel, A.,  Hoffman, Y., \& Kolatt T.,
  1997, \apj, 486, 21
  
  
\end{references}
\end{document}